\documentclass[nofootinbib,aps,prd,reprint,superscriptaddress,showkeys]{revtex4-2}

\usepackage[utf8]{inputenc} 
\usepackage{graphicx,overpic,mathtools}
\usepackage{amsthm,amsmath,amssymb,hyperref}
\usepackage{braket,bm,bbm,setspace}
\usepackage[normalem]{ulem} 
\usepackage{physics}
\usepackage{float}
\usepackage[makeroom]{cancel}
\usepackage[english]{babel}
\usepackage{xcolor}
\usepackage{tensor}
\graphicspath{ {images/} }
\addto\captionsspanish{}
\hypersetup{
	colorlinks=true,
	pdfborder={0 0 0},
	citecolor=purple,
	linkcolor=blue,
	filecolor=blue,
	urlcolor=blue,
}
\usepackage{subfigure}

\makeatletter
\def\l@subsection#1#2{}
\def\l@subsubsection#1#2{}
\makeatother

\begin{document}
\title{SQUWALS: A Szegedy QUantum WALks Simulator}
\author{Sergio A. Ortega}
%\author{Sergio A. Ortega \orcidlink{0000-0002-8237-7711}}
\email{sergioan@ucm.es}
\affiliation{Departamento de Física Teórica, Universidad Complutense de Madrid, 28040 Madrid, Spain}
\author{Miguel A. Martin-Delgado}
%\author{Miguel A. Martin-Delgado \orcidlink{0000-0003-2746-5062}}
\email{mardel@ucm.es}
\affiliation{Departamento de Física Teórica, Universidad Complutense de Madrid, 28040 Madrid, Spain}
\affiliation{CCS-Center for Computational Simulation, Universidad Politécnica de Madrid, 28660 Boadilla del Monte, Madrid, Spain.}

\begin{abstract}
	{Szegedy's quantum walk is an algorithm for quantizing a general Markov chain. It has plenty of applications such as many variants of optimizations. In order to check its properties in an error-free environment, it is important to have a classical simulator. However, the current simulation algorithms require a great deal of memory due to the particular formulation of this quantum walk. In this paper we propose a memory-saving algorithm that scales as $\mathcal{O}(N^2)$ with the size $N$ of the graph. We provide additional procedures for simulating Szegedy's quantum walk over mixed states and also the Semiclassical Szegedy walk. With these techniques we have built a classical simulator in Python called SQUWALS. We show that our simulator scales as $\mathcal{O}(N^2)$ in both time and memory resources. This package provides some high-level applications for algorithms based on Szegedy's quantum walk, as for example the quantum PageRank.}
\end{abstract}


\maketitle

\onecolumngrid

\section{Introduction}\label{Introduction}

Quantum walks are algorithms born from the quantization of classical random walks. They were first proposed in the discrete time version \cite{QRW}, and later using a continuous time \cite{Trees}. An important quantum walk in discrete time is the one introduced by Szegedy \cite{Szegedy} as a generalization of the Grover algorithm \cite{Grover}. In contrast to other approaches, which are only useful for regular graphs, this quantum walk can quantize a general Markov chain. Thus, it can be used over any arbitrary weighted graph. Moreover, it has been shown to be useful for problems of optimization \cite{Lemieux,Qfold,QMS,GWQMA}, classification \cite{Paparo1,Paparo2,APR}, quantum search \cite{Portugal,Search_walk,Searchrank,S_queries} and machine learning \cite{Paparo3}.\\

There has been research in implementing this algorithm in quantum circuits \cite{Q_circuits}. However, it is also important to have a classical form of simulating it in order to check the properties of the quantum algorithms based on this quantum walk, since these classical simulations are ideal thereby absent of errors. A naive approach for simulating it would require constructing the corresponding unitary operator. As we will see later, this grows as $\mathcal{O}(N^4)$, where $N$ is the number of nodes in the graph. However, using a sparse representation of these matrices, this dependence can be reduced to $\mathcal{O}(N^3)$. Other method based on the spectral decomposition of the operator was proposed \cite{Paparo1}, also needing memory resources scaling as $\mathcal{O}(N^3)$. Nevertheless, this scaling is prohibitive for big enough graphs as for example one with $1024$ nodes in a computer with $16$ GB of RAM memory.\\

In this work we go one step further, and propose a method for simulating this quantum walk with memory resources scaling as $\mathcal{O}(N^2)$. Furthermore, this algorithm enables the introduction of modifications as oracles \cite{Searchrank,S_queries}, which is not possible with the method based on the eigenvalues of the unitary operator, and complex phase extensions \cite{Staggered,APR}. Moreover, we show how this algorithm can be implemented to simulate the recently proposed Semiclassical Szegedy walk \cite{Semiclassical} and the quantum walk over mixed states.\\

Using this memory-saving algorithm, we have developed a Python package called SQUWALS as a simulator for Szegedy's quantum walk. There exist some simulators for quantum walks in discrete time as for example QWalk \cite{QWalk}, Hiperwalk \cite{Hiperwalk} and QuantumWalk \cite{QuantumWalk}. This last simulator implements Szegedy's quantum walk in Julia language using sparse matrices for storing the unitary operators. However, as far as we are concerned, there is not an efficient simulator for Szegedy's quantum walk that implements our algorithm. Moreover, our simulator provides some alternative formulations of the original Szegedy's quantum walk, which generalize the algorithm in a broad way.\\

This paper is structured as follows. In Section \ref{Szegedy} we review the formulation Szegedy's quantum walk. In Section \ref{Algorithm} we show the efficient algorithm for simulating this quantum walk in a classical computer. In Section \ref{Semiclassical} we review the Semiclassical Szegedy walk and show how it can be simulated with our algorithm. In section \ref{Mixed} we show an algorithm for simulating the evolution of mixed states. In section \ref{Simulator} we present SQUWALS, our Python simulator for Szegedy's quantum walk. Finally, we summarize and conclude in Section \ref{Conclusions}.

\section{Szegedy Quantum Walk}\label{Szegedy}

A classical Markov chain is a random walk that occurs in the nodes of a graph. Knowing beforehand the probability that the walker jumps from one node to another, then we can simulate the probability distribution of the walker being at each node. On one hand, for a graph with $N$ nodes, let us define the $N\times N$ classical transition matrix $G$, whose elements $G_{ji}$ are the probabilities of the walker jumping from node $i$ to node $j$. On the other hand, let $p_c(t)$ be a $N$-dimensional column vector whose elements are the probabilities of the walker being at each node at time $t$. Then, given an initial probability distribution $p_c(0)$, the classical walk can be simulated deterministically with matrix multiplications as
\begin{equation}\label{classical_evolution}
	p_c(t) = G^t p_c(0).
\end{equation}

The classical Markov chain is quantized through Szegedy's quantum walk. In this quantum walk the Hilbert space is the span of all the vectors representing the $N \times N$ directed edges of the duplicated graph, i.e.,
\begin{equation}
\mathcal{H} = \text{span}\lbrace\left|i\right>_1\left|j\right>_2,\ i,j = 0,1,...,N-1\rbrace = \mathbb{C}^N \otimes \mathbb{C}^N,
\end{equation}
where the states with indexes $1$ and $2$ refer to the nodes on two copies of the original graph. Thus, the states are defined over two quantum registers. In this paper we count the nodes of the network, and therefore the matrix indexes, from $0$ to $N-1$. We define the vectors:
\begin{equation}\label{psi_i}
	\left|\psi_i\right> := \left|i\right>_1 \otimes \sum_{k=0}^{N-1} \sqrt{G_{ki}}\left|k\right>_2,
\end{equation}
which are a superposition of the vectors representing the edges outgoing
from the $i^{th}$ vertex, whose coefficients are given by the square root of the $i^{th}$ column of the matrix $G$. From these vectors we define a projector operator onto the subspace generated by them:
\begin{equation}\label{projector}
	\Pi := \sum_{i=0}^{N-1} \left|\psi_i\right>\left<\psi_i\right|.
\end{equation}
The quantum walk operator $U$ is defined as
\begin{equation}\label{U}
	U := SR,
\end{equation}
where $R$ is a reflection over the subspace generated by the $\left|\psi_i\right>$ states,
\begin{equation}\label{reflection}
R = 2\Pi - \mathbbm{1},
\end{equation}
and $S$ is the swap operator between the two quantum registers, i.e.,
\begin{equation}\label{swap}
	S := \sum_{i,j=0}^{N-1} \left|i,j\right>\left<j,i\right|.
\end{equation}
To obtain the probability distribution of the walker after each time step of the quantum walk, the quantum state is usually projected onto the computational basis of the first register. However, the second register can also be used to measure the position instead. Let us call $_1p_q(t)$ and $_2p_q(t)$ to the vectors describing the resulting probability distribution after measuring the first or second register, respectively. Their elements are given by:
\begin{equation}
\left[\tensor[_1]{p}{_q}(t)\right]_i = \left|\left|\tensor[_1]{\left<i\right|U^{t}\left|\phi\right>}{}\right|\right|^2,
\end{equation}
\begin{equation}
\left[\tensor[_2]{p}{_q}(t)\right]_i = \left|\left|\tensor[_2]{\left<i\right|U^{t}\left|\phi\right>}{}\right|\right|^2,
\end{equation}
where $\left|\phi\right>$ is a generic initial quantum state in $\mathcal{H} = \mathbb{C}^N \otimes \mathbb{C}^N$.\\

The Szegedy's quantum walk described in this manner can be understood as a coined quantum walk where the first register encodes the position in the graph, and the second register encodes the coin state \cite{Notes}. However, the original formulation made by Szegedy consisted on a coinless quantum walk based on two reflections, each around one of the two subspaces \cite{Szegedy}. Then, in the literature we can find the following unitary operator for one step of Szegedy's quantum walk \cite{Portugal}:
\begin{equation}
W = R_BR_A,
\end{equation}
where $R_A$ is a reflection over the subspace spanned by the $\left|\psi_i\right>$ states, and $R_B$ over the space formed by the swapped version of these. Thus, we can identify $R_A = R$ and $R_B = SRS$, so that the original unitary operator $W$ corresponds to two steps of the coined version $U$, i.e., $W = U^2$. Therefore, w.l.o.g. we can use the formulation in terms of the reflection $R$ and swap $S$ operators to construct any of the unitary operator presented above, and moreover, to simulate them classically.\\

Finally, the unitary operator can be modified introducing oracles in the context of quantum search. Let us define the operator $Q$ as follows:
\begin{equation}\label{oracle}
Q = \mathbbm{1}_N - 2\sum_{k \in \mathcal{M}} \left|k\right>\left<k\right|,
\end{equation}
which inverts the sign of the vectors $\left|k\right>$ given a set $\mathcal{M}$ of nodes to mark. In order to mark the nodes in the first register, we use the oracle $Q_1$ defined as:
\begin{equation}
Q_1 = Q \otimes \mathbbm{1}_N,
\end{equation}
whereas to mark them in the second register, we use $Q_2$:
\begin{equation}
Q_2 = \mathbbm{1}_N \otimes Q.
\end{equation}
Examples of modified operators used in the literature are $U = SQ_1R$ \cite{Searchrank}, or $W = R_BR_AQ_1$ \cite{S_queries}.

\section{Memory-Saving Classical Simulation}\label{Algorithm}

In order to understand the complexity of the different simulation algorithms, we need to clarify the following concepts about the matrices:

-Dense matrix: it is a matrix where most elements are different from $0$.

-Sparse matrix: it is a matrix where most elements are $0$.\\

Moreover, let us clarify the following concepts about how a matrix can be represented in the computer:

-Dense representation: this is the naive representation, so that all the elements, including those that are $0$, are stored in memory.

-Sparse representation: in this case only the non-null elements are stored in memory. However, this method can be less efficient if the matrix is dense enough.\\

The naive method for simulating Szegedy's quantum walk is creating the unitary matrix and applying it to a vector state. The matrix representing the unitary operator $U$ in \eqref{U} is of size $N^2\times N^2$, so to simulate it classically with a dense representation we would need memory resources that scale as $\mathcal{O}(N^4)$. Nevertheless, for a dense transition matrix $G$, this unitary operator only has at most $N^3$ non-null elements. Thus, if we used a sparse representation for the unitary matrices, the resources requirements would improve to $\mathcal{O}(N^3)$. This scaling is still so much prohibitive, so that there is a limitation in simulating the quantum walk over graphs with dense transition matrices, as for example the ones used for the quantum PageRank algorithm \cite{Paparo1}. However, there are graphs with very sparse transition matrices, like regular lattices, so that the unitary operator has $\mathcal{O}(N)$ non-null elements and can be efficiently stored with a sparse representation even for thousands of nodes. Thus, this approach is feasible for sparse transition matrices, but not for dense ones.\\

Other method proposed to simulate this quantum walk was based on using the spectral decomposition of the operator \cite{Paparo1,APR}. Such a method only needs to store $2N$ eigenvectors from the dynamical subspace of the unitary operator, and since these vectors are $N^2$-dimensional, the total memory requirements scale as $\mathcal{O}(N^3)$. Thus, this algorithm is also so much prohibitive, and it is not possible to include oracles in the unitary operator.\\

Since we want an efficient algorithm for a general transition matrix $G$, w.l.o.g we are assuming that this is dense. Thus, in this section we will explain a method for simulating Szegedy's quantum walk with memory requirements scaling as $\mathcal{O}(N^2)$ for dense transition matrices. As we have seen in the previous section, any Szegedy-like quantum operator can be expressed in terms of the reflection $R$, the swap $S$ and the oracles $Q_1$ and $Q_2$. Therefore, in order to simulate this quantum walk, we provide algorithms for simulating each of the operators individually, so that they can be used as building blocks to simulate any unitary operator. Moreover, these algorithms are constructed in a vectorized manner so that we avoid the explicit use of \textit{for} loops. Thus, they are optimized for being used with the numerical Python library NumPy \cite{NumPy}, which can parallelize calculations in a CPU and is faster when the operations are vectorized.

\subsection{Matrix State Representation}

\begin{figure}[H]
	\centering
	\includegraphics[width=0.7\linewidth]{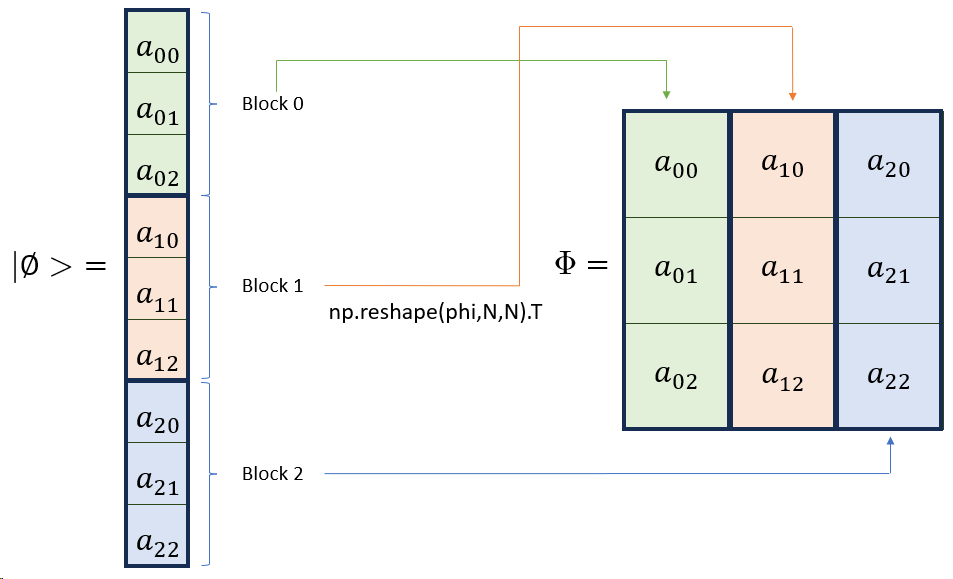}
	\caption{How to represent a vector state by a matrix state for a network with $N=3$ nodes. The blocks of the vector, indexed by the computational basis of the first register, correspond to the columns of the matrix. Beware that the indices in the coefficients $a_{ij}$ are swapped with respect to the ones of the usual matrix notation $\Phi_{ij}$.}
	\label{F:matrix}
\end{figure}

A generic vector in the doubled Hilbert space $\mathcal{H}$ can be written as
\begin{equation}\label{vector}
	\left|\phi\right> = \sum_{i,j=0}^{N-1} a_{ij} \left|i\right>_1 \left|j\right>_2.
\end{equation}
In order to simulate Szegedy's quantum walk with the least possible memory resources, we have to think of the quantum vector state as a matrix. Let us define the matrix $\Phi$ representing the vector state $\left|\phi\right>$ as the one whose elements are
\begin{equation}\label{PHI}
	\Phi_{ij} = a_{ji}.
\end{equation}
Note that using this notation, the column index represents the first register, whereas the row index represents the second register. This is so for convenience. If we divide the vector state into blocks corresponding to each state of the computational basis of the first register, then each block corresponds to each column of the matrix state. In Figure \ref{F:matrix} we show an example for a network with $N=3$ nodes.\\

\subsection{Reflection Operator}\label{S:reflection}

\subsubsection{Raw Reflection}

In order to simulate the reflection operator in \eqref{reflection}, first we need to obtain the action of the projector operator $\Pi$. The projection gives us the parallel component of the vector $\left|\phi\right>$ to the space spanned by the vectors $\left|\psi_i\right>$. Thus,
\begin{equation}
\Pi\left|\phi\right> = \left|\phi\right>_{\parallel} = \sum_{i=0}^{N-1} C_i \left|\psi_i\right>,
\end{equation}
where the coefficients $C_i$ are obtained as
\begin{equation}\label{Ci_1}
C_i = \left<\psi_i|\phi\right> = \sum_{k=0}^{N-1} a_{ik}\sqrt{G_{ki}}.
\end{equation}
Note that the only non-null elements of $\left|\psi_i\right>$ are those in the $i$-th column of $G$, so that at most only $N$ elements intervene in the dot product.\\

\begin{figure}[H]
	\centering
	\includegraphics[width=0.7\linewidth]{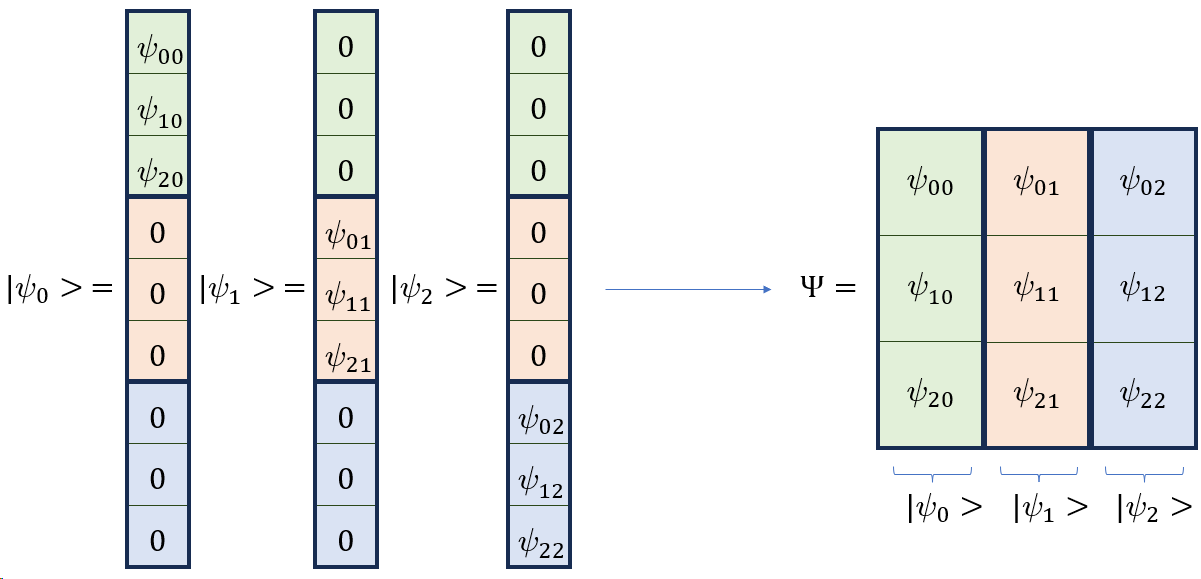}
	\caption{How to condense all the information of the $\left|\psi_i\right>$ states in a matrix for a network with $N=3$ nodes. Each column $i$ of the matrix $\Psi$ corresponds to the non-null block of each $\left|\psi_i\right>$ state.}
	\label{F:Psi}
\end{figure}

In order to calculate the coefficients $C_i$ in a vectorized way avoiding a \textit{for} loop, we create a matrix $\Psi$ where each column $i$ has the $N$ non-null elements of $\left|\psi_i\right>$. Since these correspond to the $i$-th column of the transition matrix $G$, the matrix $\Psi$ is obtained taking the element-wise square root of $G$:
\begin{equation}\label{Psi}
\Psi_{ij} = \sqrt{G_{ij}},
\end{equation}
An example of condensing the vectors $\left|\psi_i\right>$ for a graph with $N=3$ nodes is shown in Figure \ref{F:Psi}.

\begin{figure}[H]
	\centering
	\includegraphics[width=0.75\linewidth]{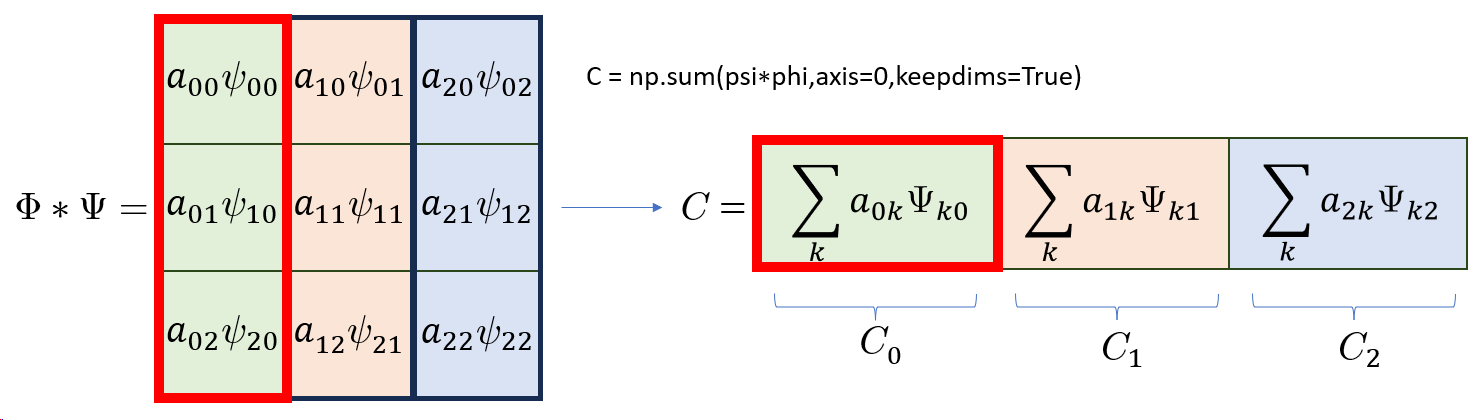}
	\caption{How to obtain the vector $C$ in \eqref{Ci_2} for a network with $N=3$ nodes. The element-wise multiplication between $\Phi$ and $\Psi$ results in a matrix whose $i$-th column has the product of the non-null elements of $\left|\psi_i\right>$ and the elements of $\left|\phi\right>$. Thus, summing the elements of each column, the coefficients $C_i$ are obtained.}
	\label{F:Ci}
\end{figure}

We can rewrite \eqref{Ci_1} as
\begin{equation}\label{Ci_2}
C_i = \sum_{k=0}^{N-1} a_{ik}\Psi_{ki}.
\end{equation}
Then, the vector $C$ with the coefficients can be obtained by multiplying element-wise the matrix state $\Phi$ with the matrix $\Psi$, and adding over the rows of the resulting matrix. In Figure \ref{F:Ci} we show an example for a network with $N=3$ nodes.\\

\begin{figure}[H]
	\centering
	\includegraphics[width=0.4\linewidth]{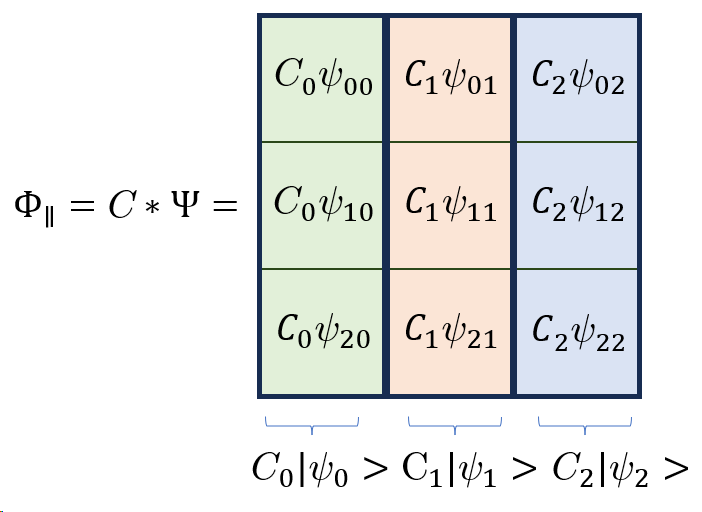}
	\caption{How to obtain the parallel component of the state $\left|\phi\right>$ for a network with $N=3$ nodes. The element-wise product of $C$ and $\Psi$ results in a matrix where each column represents the non-null elements of $C_i\left|\psi_i\right>$, which ends up representing the vector $\left|\phi_\parallel\right>$.}
	\label{F:phi_parallel}
\end{figure}

Once the coefficients are obtained, we can use the broadcasting feature of NumPy to multiply element-wise the row vector $C$ with the matrix $\Psi$. The resulting matrix has in the $i$-th column the non-null elements of $\left|\psi_i\right>$ multiplied by the coefficient $C_i$. Thus, it results in the matrix representing the parallel component of our state, $\left|\phi\right>_\parallel$. An example is shown in Figure \ref{F:phi_parallel}.

Finally, the reflection of the state is obtained straightforwardly as
\begin{equation}\label{subtraction}
R\left|\psi\right> = 2\left|\psi\right>_\parallel - \left|\psi\right>.
\end{equation}

So far we have devised an algorithm to simulate the original reflection operator. However, it can be extended introducing complex phases. Next, we will show two different forms of extending this operator and how to implement it in the simulation.

\subsubsection{Extended Szegedy's Quantum Walk}

The extended Szegedy's Quantum Walk was introduced in \cite{Staggered}. It consist on modifying the $\left|\psi_i\right>$ vector with complex phases as follows:
\begin{equation}\label{psi_i_phases}
	\left|\psi_i\right> := \sum_{k=0}^{N-1} e^{i\theta_{ik}}\sqrt{G_{ki}}\left|i\right>_1\left|k\right>_2,
\end{equation}
In order to introduce this modification in our algorithm, we have to modify the matrix $\Psi$ in \eqref{Psi} by
\begin{equation}
	\Psi_{ij} = e^{i\theta_{ji}}\sqrt{G_{ij}},
\end{equation}
which can be done easily creating a matrix $\Theta$ with the complex phases, and multiplying its transpose element-wise with the element-wise square root of the transition matrix $G$. The original reflection operator is recovered when all the phases $\theta_{ij}$ are equal to $0$.

\subsubsection{Arbitrary Phase Rotations}

The other modification consist on transforming the reflection operator into an arbitrary phase rotation \cite{APR} as
\begin{equation}
R(\theta) = (1-e^{i\theta})\Pi - \mathbbm{1}.
\end{equation}
To simulate it we just have to modify expression \eqref{subtraction} as
\begin{equation}
	R(\theta)\left|\psi\right> = (1-e^{i\theta})\left|\psi\right>_\parallel - \left|\psi\right>,
\end{equation}
in the last step of the algorithm. The original reflection is recovered when the phase $\theta$ is equal to $\pi$.

\subsection{Swap Operator}

The action of the swap operator over a vector of the doubled computational basis is
\begin{equation}
S\left|i\right>_1\left|j\right>_2 = \left|j\right>_1\left|i\right>_2.
\end{equation}
Thus, it corresponds to exchange the coefficients $a_{ij}$ of the vector state \eqref{vector} with the coefficients $a_{ji}$. Given the matrix form of expressing the vector state, this operation corresponds straightforwardly to the transposition of the matrix $\Phi$.

\subsection{Oracle Operators}

The action of the oracle operator in \eqref{oracle} over a state of the computational basis is
\begin{equation}
	Q\left|i\right> := 
	\left\lbrace\begin{array}{c}
		-\left|i\right> \ \ \ \ \ \text{if} \ i \in \mathcal{M},\\
		\ \ \ \ \left|i\right> \ \ \ \ \    \text{otherwise},
	\end{array}
	\right.
\end{equation}
where $\mathcal{M}$ is a set of marked nodes.\\

\begin{figure}[H]
	\centering
	\includegraphics[width=0.7\linewidth]{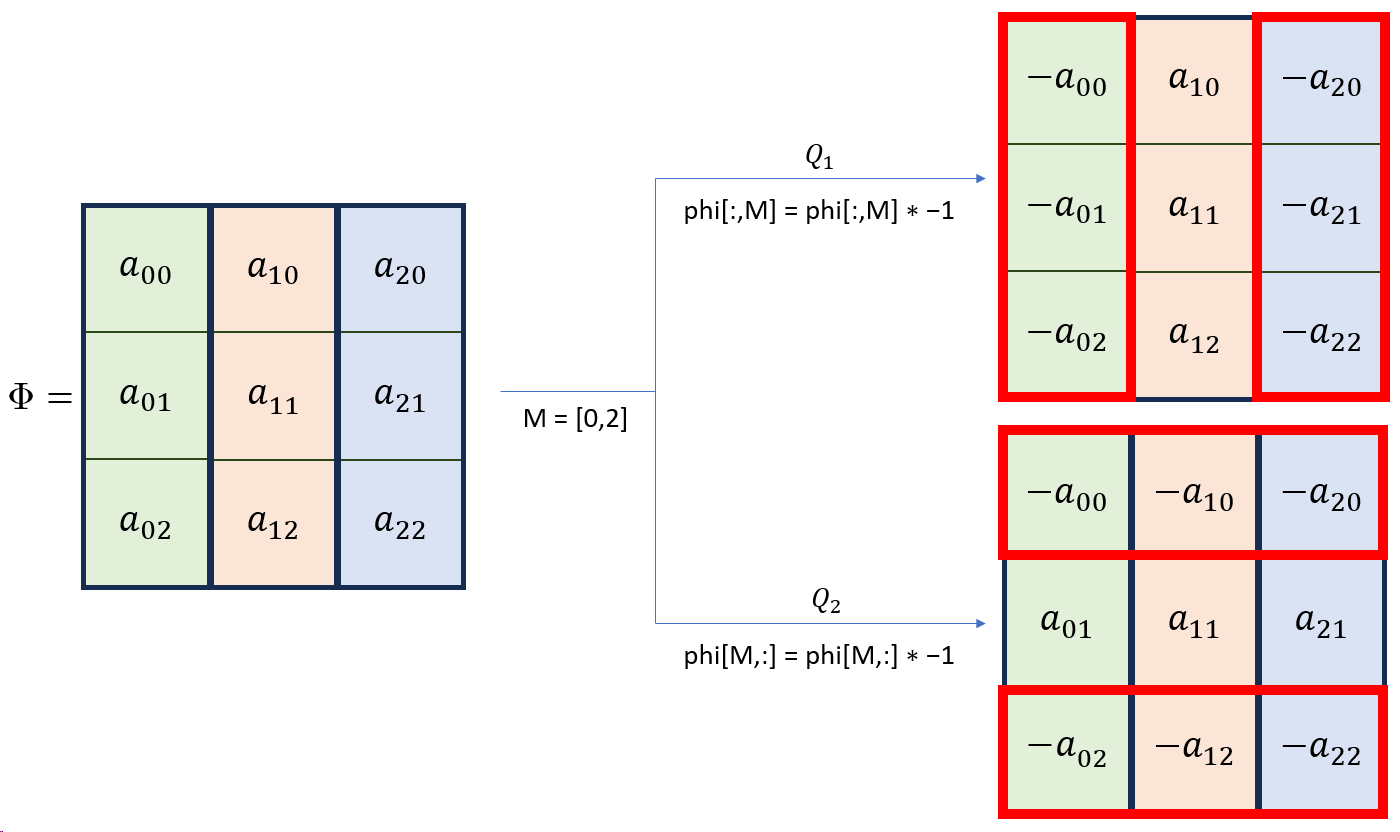}
	\caption{How to apply the oracle operators for a network with $N=3$ nodes. In this case nodes $0$ and $2$ are being marked. In order to mark them in the first (second) register, the corresponding columns (rows) are multiplied by $-1$.}
	\label{F:marked}
\end{figure}

On one hand, in order to mark node $i$ in the first register, we have to invert the sign of the coefficients $a_{ij} \ \forall \ j$ of the vector state in \eqref{vector}. Given the matrix form $\Phi$ representing this vector, the index $i$ in $a_{ij}$ refers to the columns. Thus, we have to invert the sign of the elements along the $i$-th column. Having a list of marked nodes, we can multiply the columns corresponding to the marked nodes in a vectorized form in Python. On the other hand, if we wanted to mark the nodes in the second register, we could proceed in a similar way but inverting the sing of the elements along the corresponding rows. An example is shown in Figure \ref{F:marked} for a network with $N=3$ nodes, where nodes $0$ and $2$ are being marked.

The same as the reflection operator, the oracle can be extended with an arbitrary phase rotation as
\begin{equation}\label{oracle_phase}
	Q(\theta) = \mathbbm{1}_N - (1-e^{i\theta})\sum_{k \in M} \left|k\right>\left<k\right|,
\end{equation}
whose action over the computational basis is
\begin{equation}
	Q(\theta)\left|i\right> := 
	\left\lbrace\begin{array}{c}
		e^{i\theta}\left|i\right> \ \ \ \ \ \text{if} \ i \in \mathcal{M},\\
		\ \ \ \ \ \ \left|i\right> \ \ \ \ \    \text{otherwise}.
	\end{array}
	\right.
\end{equation}
Thus, to introduce the phase in the algorithm we just have to multiply the elements by $e^{i\theta}$ instead of by $-1$.

\subsection{Measurement}

The probability of measuring node $i$ after measuring the first register is given by:
\begin{equation}\label{probability_1}
	\left[\tensor[_1]{p}{_q}\right]_i = ||\tensor[_1]{\left<i|\phi\right>}{}||^2 = \left|\left|\sum_{k,l=0}^{N-1} a_{kl} \ \tensor[_1]{\left<i|k\right>}{_1} \left|l\right>_2\right|\right|^2 = \left|\left|\sum_{l=0}^{N-1} a_{il} \left|l\right>_2\right|\right|^2 = \sum_{l=0}^{N-1} |a_{il}|^2,
\end{equation}
where we have used in the last step that the computational basis is ortonormal.
Analogously, the probability of measuring node $i$ if we were measuring the second register is given by:
\begin{equation}\label{probability_2}
	\left[\tensor[_2]{p}{_q}\right]_i = ||\tensor[_2]{\left<i|\phi\right>}{}||^2 = \left|\left|\sum_{k,l=0}^{N-1} a_{kl} \left|k\right>_1  \tensor[_2]{\left<i|l\right>}{_2}\right|\right|^2 = \left|\left|\sum_{k=0}^{N-1} a_{ki} \left|l\right>_2\right|\right|^2 = \sum_{k=0}^{N-1} |a_{ki}|^2.
\end{equation}

\begin{figure}[H]
	\centering
	\includegraphics[width=0.8\linewidth]{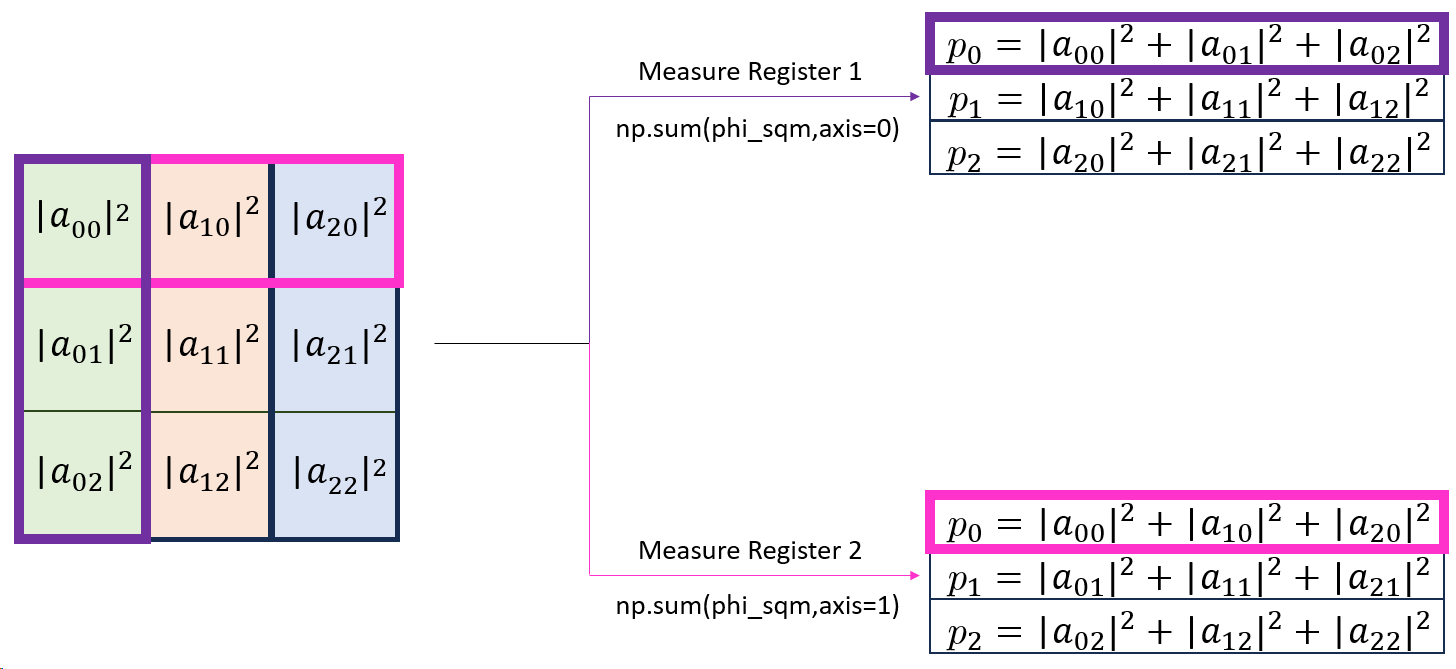}
	\caption{How to obtain the probability distributions of the walker for a network with $N=3$ nodes. First, the square modulus of all the elements in $\Phi$ is taken. To obtain the probabilities after measuring in the first (second) register, the elements of each column (row) are summed.}
	\label{F:measure}
\end{figure}

To obtain the probabilities of measuring each of the nodes, first we need to take the square modulus of all the elements in the matrix state $\Phi$. After that, the probability of measuring node $i$ in the first (second) register is obtained by adding the elements of the $i$-th column (row), according to \eqref{PHI}, \eqref{probability_1} and \eqref{probability_2}. A vectorized way of doing this with NumPy is shown in Figure \ref{F:measure}.\\

\subsection{Initial State}

In many applications of Szegedy's quantum walk, as for example quantum search \cite{S_queries} and quantum PageRank \cite{Paparo1}, the initial state of the walker is an equal superposition of the $\left|\psi_i\right>$ states. Let us call $\left|\Psi^{(0)}\right>$ to this initial state, so that
\begin{equation}\label{initial}
\left|\Psi^{(0)}\right> = \frac{1}{\sqrt{N}}\sum_{i=0}^{N-1} \left|\psi_i\right>.
\end{equation}
It can be seen that the matrix $\Psi$ corresponds to the sum of the matrices representing the $\left|\psi_i\right>$ states. Thus, the matrix representing this initial state is obtained just dividing the matrix $\Psi$ by $\sqrt{N}$.

\subsection{Memory and Time Complexity}\label{S:MTC}

Given the the procedures described for simulating the action of the different operators, we can formulate the following lemmas about our algorithm.\\

{\bfseries{Lemma 1. Memory: The memory storage of SQUWALS scales as $\mathcal{O}(N^2)$.}} In all the algorithms that we have seen in this section, the objects are dense $N\times N$ matrices. Thus, the number of elements that we have to store for each application of the unitary operator scales as $\mathcal{O}(N^2)$. For a dense classical matrix we just have $N^2$ elements, so that this scaling of the memory requirements is the minimum achievable, and then it is optimal. Moreover, since we do not need the quantum states at each time step, but the probabilities, we can just measure each time we apply the algorithm for simulating the unitary evolution without saving the intermediate quantum states. Note that at each step we have $N$ probability elements, so for big enough networks the memory needed to store them is negligible with respect to the cost of storing the quantum state. Therefore, the memory requirements would not increase with the number of time steps, and only depend on the size of the graph as $\mathcal{O}(N^2)$.\\

{\bfseries{Lemma 2. Time Complexity: The time complexity of the SQUWALS algorithm scales as $\mathcal{O}(N^2)$.}} Regarding the time complexity of the algorithm, the most costly operations are the multiplications. As we have seen, all the multiplications are done element-wise, so that each element of the $N\times N$ matrices intervene only once for each of the operations. Thus, the number of times that an element is used does not depend on $N$. Since in each element-wise multiplication we have $N^2$ elements intervening, then the time needed by the algorithm to run will scale theoretically as $\mathcal{O}(N^2)$, the same as the memory requirements. Finally, since we need to calculate all the time steps sequentially, the time required by the algorithm will scale linearly with the number of time steps.\\

The discussion above is valid for dense representation of the matrices, where even null elements are being stored. However, if the transition matrix only has very few non-null elements, then we could use a sparse representation to reduce the requirements of our algorithm even further. For example, for a cyclic 2D lattice the number of non-null elements in $G$ is only $4N$, since there are $N$ nodes with $4$ connections each.\\

\section{Classical Simulation of Semiclassical Szegedy Walk}\label{Semiclassical}

Recently, a new kind of walk algorithm that mixes quantum and classical properties has been proposed, called Semiclassical Szegedy walk \cite{Semiclassical}. From a functional point of view in a quantum computer, this walk consist on repeated measurements of the walker position at regular intervals of time. There are two parameters to describe a semiclassical walk. The first one is the quantum time \cite{QT} $t_q$, which is the number of times we apply the unitary evolution between measurements. The second one is the classical time $t_c$, which is the number of times that the scheme of quantum evolution and measurement is repeated.\\

From a formal point of view, a semiclassical walk can be understood as a classical walk where the transition matrix encodes the quantum evolution. The application of this semiclassical transition matrix corresponds to a classical step of the semiclassical walk. Thus, we actually have an entire family of semiclassical walks, parameterized by the quantum time $t_q$, and the classical time is the actual parameter that determines the time steps of the walker. Moreover, depending of which register is being measured, we can have different classes of semiclassical walks. Class I corresponds to those where the first register is measured, whereas class II corresponds to those where the second one is measured instead.\\

An intrinsic feature of the semiclassical walks is that after the measurement of the walker position, the quantum system is reset so that if the result of the measurement is node $i$, then $\left|\psi_i\right>$ is prepared. Thus, to simulate them deterministically in a classical computer we need the semiclassical matrices defined as follows:
\begin{equation}\label{G1}
	\tensor[_1]{G}{}^{(t_q)}_{ji} := \left|\left|\tensor[_1]{\left<j\right|U^{t_q}\left|\psi_i\right>}{}\right|\right|^2,
\end{equation}
\begin{equation}\label{G2}
	\tensor[_2]{G}{}^{(t_q)}_{ji} := \left|\left|\tensor[_2]{\left<j\right|U^{t_q}\left|\psi_i\right>}{}\right|\right|^2,
\end{equation}
where the left-subscript in the semiclassical matrices denotes the class of the walk, and the quantum time $t_q$ characterizes the particular semiclassical walk of the family.\\

Since a classical walk is easy to simulate given a transition matrix, the semiclassical walk simulation reduces to calculate these semiclassical matrices. For this, we have to simulate the quantum evolution of all the $\left|\psi_i\right>$ states. A naive approach would be to use our simulation algorithm $N$ times with a \textit{for} loop, taking each of the $\left|\psi_i\right>$ states as initial states. In this case, the memory resources would still scale as $\mathcal{O}(N^2)$ since only one quantum evolution is being performed at a time, and the time needed would multiply by $N$, scaling as $\mathcal{O}(N^3)$.\\

\begin{figure}[H]
	\centering
	\includegraphics[width=0.6\linewidth]{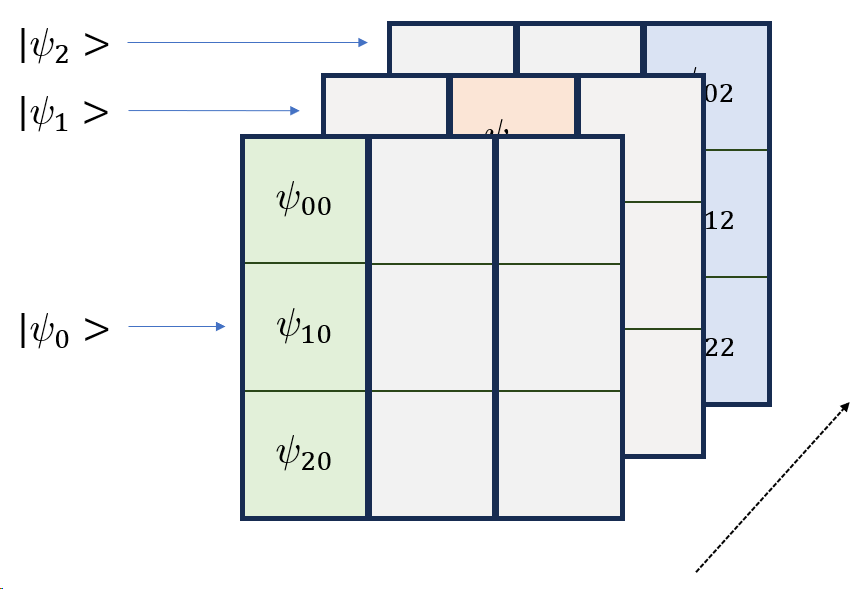}
	\caption{How to stack different initial states for a network with $N=3$ nodes. The different matrix states are stacked along a third dimension, forming a 3D tensor. In this case the three $\left|\psi_i\right>$ states are stacked in a batch.}
	\label{F:tensor}
\end{figure}

If the memory resources of the computer are permissive enough, we can save time vectorizing the operations with NumPy. We can stack different initial matrix states along a third dimension, so that the broadcasting feature allows to perform the quantum evolution over all of them with the same operations that we have seen in the previous section. An example of a vectorized initial state with all the $\left|\psi_i\right>$ states is shown in Figure \ref{F:tensor} for a network with $N=3$ nodes.\\

In the case of vectorizing all the operations, the memory requirements would scale now as $\mathcal{O}(N^3)$. Note that the time complexity remains as $\mathcal{O}(N^3)$ since it only depends on the number of operations, although these can be done faster thanks to the vectorization and how they could be parallelized. Depending on the memory resources of the computer, we could make a batch with all the $\left|\psi_i\right>$ states, or use mini-batches composed of a limited number of states to save memory. In the limiting case of one state at a time the scaling of the memory as $\mathcal{O}(N^2)$ is recovered.

\section{Classical Simulation with Mixed States}\label{Mixed}

Since the Hilbert space $\mathcal{H}$ is of dimension $N^2$, a naive approach for simulating the evolution of a mixed state would require to store an $N^2\times N^2$ matrix, which is prohibitive and would need memory resources scaling as $\mathcal{O}(N^4)$. Moreover, depending on the state it can also be prohibitive if we used a sparse representation of this matrix. In this section we show a method for simulating Szegedy's quantum walk over a mixed state that is expressed as a weighted sum of a particular ortonormal set of pure states. In contrast to the simulation algorithm over pure states, which can be used to obtain all the complex amplitudes of the state, the algorithm that we provide for mixed states does not calculate the matrix elements of the density operator, but only the probabilities of measuring each of the nodes.\\

Let us denote $\left|b_i\right>$ to the vectors of an ortonormal set that are in the doubled Hilbert space $\mathcal{H}$. The mixed state is then expressed as
\begin{equation}\label{mixed}
\rho = \sum_{i=0}^{M-1} c_i \left|b_i\right>\left<b_i\right|,
\end{equation}
where $c_i$ are non-negative real numbers adding 1, and $M$ is the size of the set. Note that this size does not have to be $N^2$ in principle.\\

For a given generic unitary evolution $U$, the mixed state after the evolution is:
\begin{equation}
U \rho U^\dagger = \sum_{i=0}^{M-1} c_i \ U \left|b_i\right>\left<b_i\right| U^\dagger.
\end{equation}

Let us suppose that we want to measure the first register of the system. The probability of measuring node $j$ is given by:
\begin{equation}\label{trace_1}
\left[\tensor[_1]{p}{_q}\right]_j = \text{Tr}_2 \left[\tensor[_1]{\left<j\right|U \rho U^\dagger\left|j\right>}{_1}\right] = \text{Tr}_2\left[\sum_{i=0}^{M-1} c_i \tensor[_1]{\left<j\right|U \left|b_i\right>\left<b_i\right| U^\dagger\left|j\right>}{_1}\right].
\end{equation}
The same as the product of a vector state with a vector of the computational basis of the first register results in a vector in the second subspace, in this case the sandwich results in a density matrix in the second subspace. Thus, we have to trace over the second subspace to obtain the final probability. Since the product $\tensor[_1]{\left<j\right|U \left|b_i\right>}{}$ results in a vector in the second register, we can express the trace as:
\begin{equation}\label{trace_2}
\text{Tr}_2 \left[\sum_{i=0}^{M-1} c_i \tensor[_1]{\left<j\right|U \left|b_i\right>\left<b_i\right| U^\dagger\left|j\right>}{_1}\right] = \sum_{i=0}^{M-1} c_i \left|\left|\tensor[_1]{\left<j\right|U \left|b_i\right>}{}\right|\right|^2.
\end{equation}
The term $\left|\left|\tensor[_1]{\left<j\right|U \left|b_i\right>}{}\right|\right|^2$ corresponds to the probability of measuring node $j$ after the quantum walk evolution of the state $\left|b_i\right>$. Thus, the probability of measuring node $j$ starting from the mixed state is a weighted mean of the results for each of the states that form the mixed state, where the weights are given by the coefficients $c_i$.\\

If we were measuring the second register instead, a similar result would be obtained. Thus, in order to simulate the probabilities of the walker from a mixed state we need to simulate the quantum walk for each of the states $\left|b_i\right>$ that comprise it. The same as in the semiclassical walk simulation, we can stack them in a batch to simulate them in a vectorized form, as long as the memory resources allow it. However, we can simulate one at a time, so that the memory requirements again scale as $\mathcal{O}(N^2)$.

\subsection{Relation with the Semiclassical Walk}

We have previously mentioned that the common initial state for Szegedy's quantum walk in \eqref{initial} is formed by a superposition of the $\left|\psi_i\right>$ states. Thus, these states could be an appropriate set to form an initial mixed state as
\begin{equation}\label{mix_psi}
	\rho = \sum_{i=0}^{N-1} c_i \left|\psi_i\right>\left<\psi_i\right|.
\end{equation}

Generalizing the evolution for a quantum time step $t_q$ and using \eqref{trace_1}, \eqref{trace_2} and \eqref{G1}, we obtain a relation between the probabilities measuring the first register and the semiclassical matrices of class I:
\begin{equation}
\left[\tensor[_1]{p}{_q}\right]_j = \text{Tr}_2 \left[\tensor[_1]{\left<j\right|U^{t_q} \rho U^{t_q\dagger}\left|j\right>}{_1}\right] = \sum_{i=0}^{N-1} \tensor[_1]{G}{}^{(t_q)}_{ji} c_i.
\end{equation}
If $c_i$ are the coefficients of a column probability vector, then the quantum walk over the mixed state in \eqref{mix_psi} during $t_q$ steps corresponds to a classical step of the corresponding semiclassical walk over the probability distribution given by the coefficients $c_i$. The same discussion is valid for measuring in the second register, using the semiclassical walks of class II. From the point of view of the simulation in a classical computer, this means that the simulation of the semiclassical walks by means of the semiclassical matrices implies solving the problem for any mixed state of the form in \eqref{mix_psi}.\\

\section{SQUWALS}\label{Simulator}

Given the algorithms presented above, we have developed a Python library called SQUWALS, which performs the operations in a vectorized form using NumPy. This software allows the user to simulate Szegedy's quantum walk in a simple manner without a deep knowledge of the algorithm itself. Moreover, the user does not have to transform the quantum states into matrices, so that they can be provided just as normal NumPy vectors.\\

In this section we describe the code fundamentals for using our simulator, and show that the scaling in both the time and memory needed to run it is compatible with $\mathcal{O}(N^2)$ as theoretically predicted. Finally, we show some applications based on Szegedy's quantum walk that are available in our library.

\subsection{Simulator Fundamentals}

In order to simulate a Szegedy's quantum walk, the user needs to provide an initial state vector $\left|\phi\right>$ and a classical transition matrix $G$. This matrix must be column-stochastic, so that each column sums up to 1. In case that the user wants to use the initial quantum superposition state in \eqref{initial}, we provide a function that creates it from the transition matrix.\\

Our library provides three classes for the three main operators, which are the reflection $R$, the swap $S$ and the oracle $Q$. On one hand, the reflections are created providing the transition matrix $G$, and optionally, the phase extensions mentioned in \ref{S:reflection}. On the other hand, the oracle is created providing a list with the marked nodes, the subspace where the marking is having place, and an optional phase. Once the necessary operators have been instantiated, the user must use the unitary class to create an unitary operator concatenating the building-block operators in the desired manner. The unitary object can then be fed with a quantum state vector, and returns the result of the quantum evolution. After that, in order to obtain the probability distributions, the measurement class must be used, choosing what register is being measured.\\

With the unitary and measurement classes we can simulate Szegedy's quantum walk from any initial quantum state. However, this would result in a low-level implementation, since the user must code a loop for the different time steps and manage to store the results properly. For that reason, we provide a higher-level simulator that does it for us. Our quantum simulator needs as input parameters the unitary operator $U$, as for example the one in \eqref{U}, and the initial state $\left|\phi\right>$, as well as the number of time steps and the register to measure. This simulator performs all the time steps while measuring the probabilities, and it only provides the probability distributions for all time steps. Thus, the intermediate quantum states are not being stored and the memory requirements do not scale with the number of steps.\\

Finally, in case we want to simulate the quantum evolution over different initial quantum states, our simulator allows the user to introduce a batch of quantum states, so that their evolutions are done in a vectorized form. An example is the case of calculating the evolution over a mixed state, where we need to simulate the evolution of all the states that comprise it. Moreover, once the simulation over this set of states is performed, the result can be fed to another utility that receives the coefficients in \eqref{mixed} and returns the probability distributions for the evolution of the mixed state.\\

\subsection{Time and Memory Complexity}

In \ref{S:MTC} we proved that the theoretical scaling of both the memory requirements and time needed for the SQUWALS algorithm to run is as $\mathcal{O}(N^2)$. In order to check that our implementation follows that scaling, we have made some simulations for random dense transition matrices of growing sizes from $N=100$ to $N=16000$. We have used an AMD Ryzen 9 5950X 16-Core Processor for all the simulations.\\

\begin{figure}[H]
	\centering
	\subfigure[]{\includegraphics[scale=0.5846]{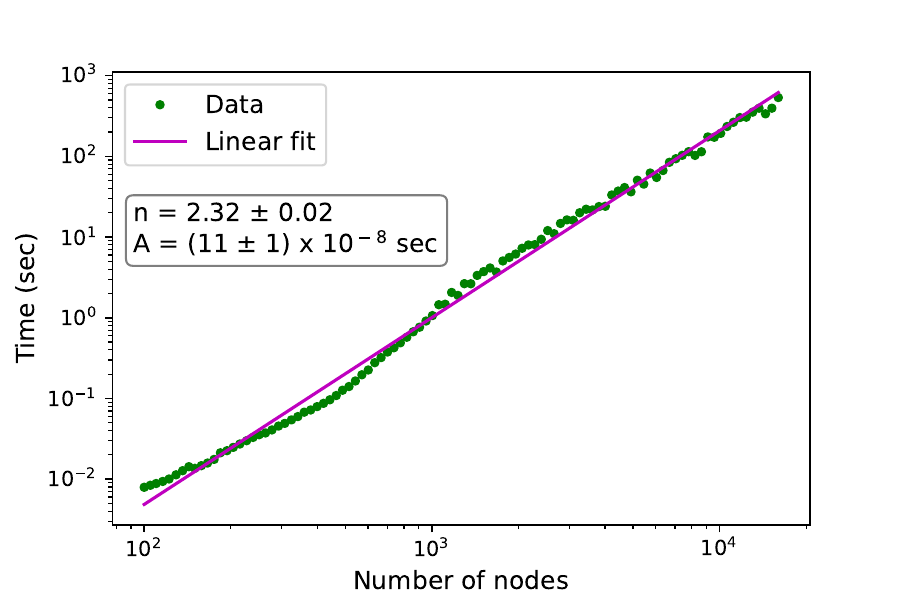}\label{F:time_all}}
	\subfigure[]{\includegraphics[scale=0.5846]{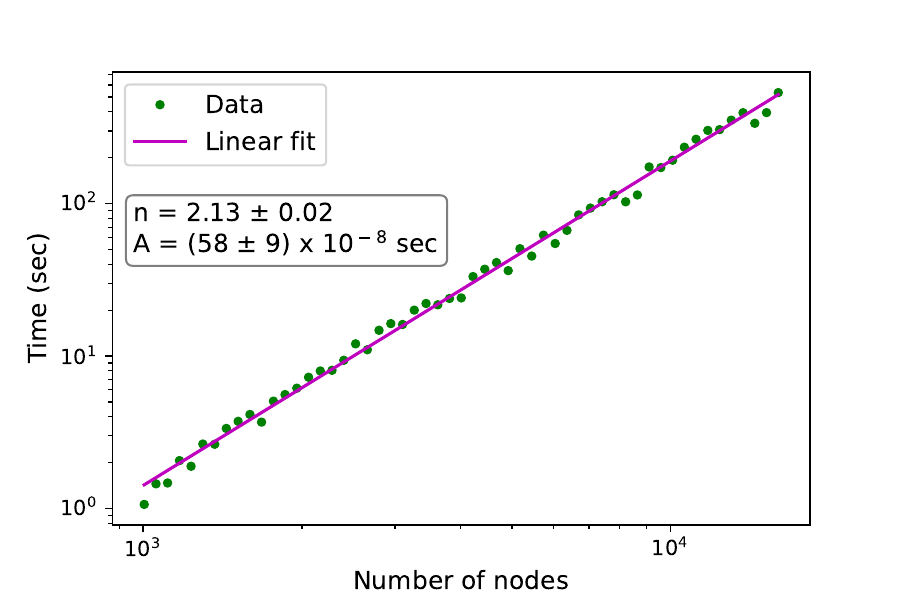}\label{F:time_region}}
	\caption{Linear fit in logarithmic scale of the time needed to run the algorithm versus the number of nodes of the graph. a) From $N=100$ to $N=16000$. b) From $N=1000$ to $N=16000$. All the simulations have been performed for 100 time steps, using the double unitary operator $W=SRSR$. The errors are given by the standard deviations provided by the fitting.}
	\label{F:time}
\end{figure}

First, we have measured the time it takes to run the algorithm. We want to fit the data to the following scaling law:
\begin{equation}\label{fit}
\text{Time} = AN^n,
\end{equation}
so that the coefficient $n$ determines the scaling complexity. To do so, we have taken logarithms in \eqref{fit} so that
\begin{equation}
\log\left(\text{Time}\right) = \log A + n\log N,
\end{equation}
and made a linear fit in logarithmic scale. The results are shown in Figure \ref{F:time_all}. As we can see, the exponent $n$ is significantly greater than $2$. However, note that the linear behavior starts from $N = 1000$ approximately. Recall that the theoretical scaling was discussed in the asymptotic limit of large $N$. This can explain this behavior, and for these reason we have made the fit in the linear region from $N=1000$ in Figure \ref{F:time_region}. In this case now we can see a clear linear trend in logarithmic scale, with a smaller exponent of $n = 2.13$. Although this is slightly larger than $2$, it seems that our implementation has a time complexity approximately similar to the theoretical one of $\mathcal{O}(N^2)$.

\begin{figure}[H]
	\centering
	\subfigure[]{\includegraphics[scale=0.5846]{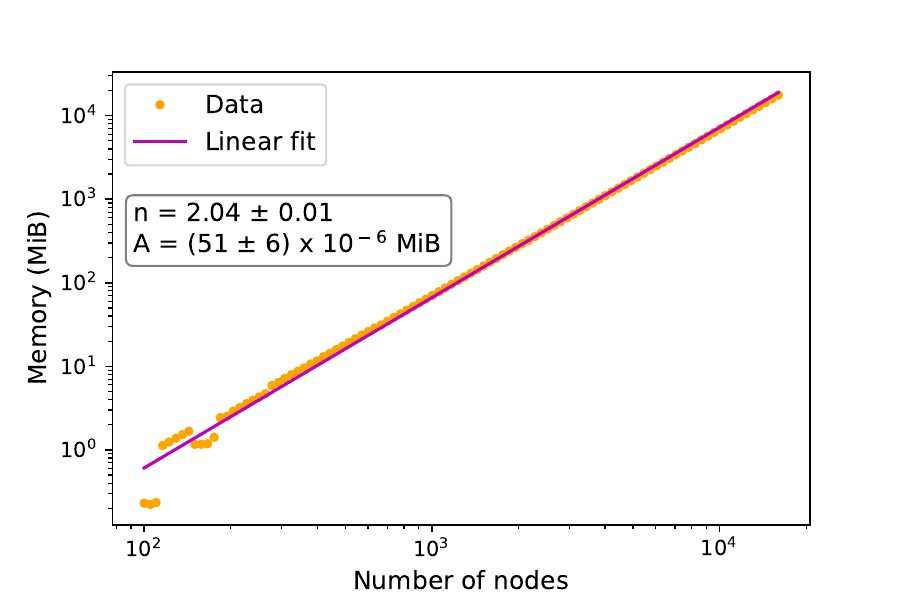}\label{F:memory_all}}
	\subfigure[]{\includegraphics[scale=0.5846]{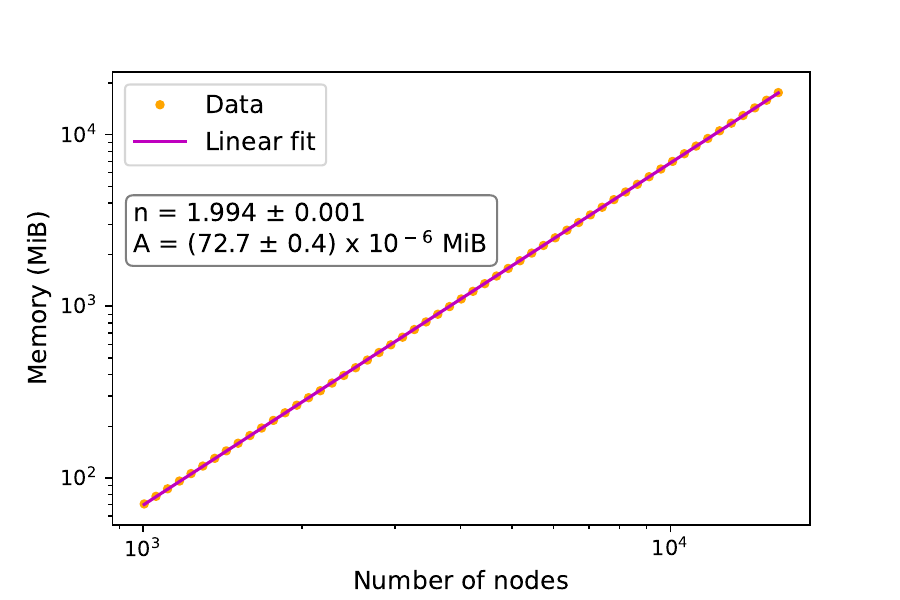}\label{F:memory_region}}
	\caption{Linear fit in logarithmic scale of the memory consumed by the algorithm versus the number of nodes of the graph. a) From $N=100$ to $N=16000$. b) From $N=1000$ to $N=16000$. All the simulations have been performed for 100 time steps, using the double unitary operator $W=SRSR$. The errors are given by the standard deviations provided by the fitting.}
	\label{F:memory}
\end{figure}

For the scaling of the memory requirements, we have follow a similar procedure measuring the memory consumed by the algorithm. The results are shown ins Figure \ref{F:memory_all}. In this case we have again finite size effects for small values of $N$. Thus, in Figure \ref{F:memory_region} we show the fit for larger values of $N$, where the asymptotic behavior occurs. The exponent $n \approx 2$ in this case, so that our simulator has the expected theoretical scaling with the memory resources as $\mathcal{O}(N^2)$. 

\subsection{High-Level Applications}

With the main simulator described above we can simulate any algorithm based on Szegedy's quantum walk. Nevertheless, we aim to provide higher-level functionalities in our library so that end users can apply different Szegedy-based algorithms easier.\\

An application implemented in our library is the semiclassical Szegedy walk. We provide a function that manages to create the initial $\left|\psi_i\right>$ states from a transition matrix $G$ and simulate them using an unitary operator $U$, to return the semiclassical matrices in $\eqref{G1}$ and $\eqref{G2}$ directly. Moreover, this function accepts a batch size parameter, so that the user can define the size of the batch of states that are vectorized at a time. Later, to simulate the walk, we also provide a classical simulator, so that it calculates a semiclassical walk using the desired semiclassical transition matrix $G$.\\

Another application is the quantum PageRank algorithm described in \cite{Paparo1,Paparo2}. This is valuable not only as a quantum algorithm implementable in a quantum computer, but also as a quantum inspired classical algorithm thought to be run in a classical computer. Moreover, the transition matrix has no null elements. For these reasons, having an efficient classical implementation is crucial. In this case, the algorithm starts from the initial quantum superposition in \eqref{initial}, uses the double Szegedy unitary $U^2 = SRSR$ and measures in the second register. After that, the probability distributions are averaged in time to yield the quantum PageRank distribution. Our function only requires the classical transition matrix to perform all the calculations. Moreover, it allows the input of two arbitrary phase rotations as introduced in \cite{APR}.\\

\section{Conclusions}\label{Conclusions}

We have reviewed the formulation of Szegedy's quantum walk in a general form, so that any unitary operator can be decomposed on three building blocks: reflections, swaps and oracles. These operators are of size $N^2 \times N^2$, where $N$ is the number of nodes of the graph, so that a naive approach for simulating this quantum walk would need memory resources that scale as $\mathcal{O}(N^4)$. There exist other methods that only requires $\mathcal{O}(N^3)$ memory resources when the transition matrix is dense, as the use of a sparse representation for the unitary operators. For sparse transition matrices this method may require so much less memory resources that is able to simulate networks with thousand of nodes. However, for a general transition matrix this scaling is so much prohibitive for big networks as for example $1024$ nodes.\\

In order to save memory resources even if the transition matrix is dense, we have devised an algorithm that reduces the memory and time requirements further to $\mathcal{O}(N^2)$, and allows several modifications of Szegedy's unitary evolution as the introduction of oracles and complex phases extensions.\\

We have also reviewed the semiclassical Szegedy walk and devised a method for simulating the quantum walk over mixed states. Both algorithms require simulating the walk over a set of different initial states, and we have shown how that procedure can be vectorized constructing batches of quantum states. So far, for simulating over mixed states our algorithm requires that they are diagonal in an orthogonal set, so that there are no coherence matrix elements. This is an open problem that needs a further research in the future.\\

Finally, we have presented SQUWALS, a Python package for simulating Szegedy's quantum walk using our memory-saving algorithm. This is a NumPy based simulator that leverages vectorization techniques to speed up the calculations. We have coded a high-level functionality that only requires the initial state and the transition matrix to be provided by the user, and performs all the needed calculations to obtain the probability distributions at each time step of the quantum walk. Moreover, we have shown that both the time and memory requirements of our simulator scale as $\mathcal{O}(N^2)$, as theoretically predicted.\\

In the future, we expect our algorithm to be parallelized with specialized hardware for that task as for example GPU cards. Moreover, we aim to continue adding more higher-level algorithms to our library, having as yet algorithms for quantum PageRank and semiclassical walks simulation.\\

\section{Data Availability Statement}\label{Data}

The simulator library is on GitHub: \url{https://github.com/OrtegaSA/SQUWALS-repo}.\\

There is a tutorial for using the simulator, and two tutorials for the two high-level applications.

\section*{Acknowledgments}
We acknowledge the support from the CAM/FEDER Project No.S2018/TCS-4342 (QUITEMAD-CM), Spanish MINECO grants MINECO/FEDER Projects,  PID2021-122547NB-I00 FIS2021, the “MADQuantum-CM" project funded by Comunidad de Madrid and by the Recovery, Transformation, and Resilience Plan – Funded by the European Union - NextGenerationEU and Ministry of Economic Affairs Quantum ENIA project. M. A. M.-D. has been partially supported by the U.S.Army Research Office through Grant No. W911NF-14-1-0103. S.A.O. acknowledges support from Universidad Complutense de Madrid - Banco Santander through Grant No. CT58/21-CT59/21.

\bibliography{MiBiblio}

\begin{thebibliography}{10}
\bibitem{QT}
{By quantum time we refer to a parameter that determines the duration of the
  quantum evolution, rather than a quantum operator.}
\end{thebibliography}
\bibliographystyle{unsrt}

\end{document}